\documentclass[conference]{IEEEtran}
\IEEEoverridecommandlockouts
\usepackage{amsmath,amssymb,amsfonts}
\usepackage[linesnumbered,ruled,vlined]{algorithm2e}
\usepackage{algpseudocode}
\usepackage{graphicx}
\usepackage{textcomp}
\usepackage{xcolor}
\usepackage{url}
\begin{document}

\title{Laparoscopic Image Desmoking Using the U-Net with New Loss Function and Integrated Differentiable Wiener Filter}

\author{\IEEEauthorblockN{1\textsuperscript{st} Chengyu Yang}
\IEEEauthorblockA{\textit{Department of Computer Science} \\
\textit{New Jersey Institute of Technology}\\
Newark, New Jersey, USA \\
cy322@njit.edu}
\and
\IEEEauthorblockN{2\textsuperscript{nd} Chengjun Liu}
\IEEEauthorblockA{\textit{Department of Computer Science} \\
\textit{New Jersey Institute of Technology}\\
Newark, New Jersey, USA \\
chengjun.liu@njit.edu}
% \and
% \IEEEauthorblockN{3\textsuperscript{rd} Given Name Surname}
% \IEEEauthorblockA{\textit{dept. name of organization (of Aff.)} \\
% \textit{name of organization (of Aff.)}\\
% City, Country \\
% email address or ORCID}
% \and
% \IEEEauthorblockN{4\textsuperscript{th} Given Name Surname}
% \IEEEauthorblockA{\textit{dept. name of organization (of Aff.)} \\
% \textit{name of organization (of Aff.)}\\
% City, Country \\
% email address or ORCID}
% \and
% \IEEEauthorblockN{5\textsuperscript{th} Given Name Surname}
% \IEEEauthorblockA{\textit{dept. name of organization (of Aff.)} \\
% \textit{name of organization (of Aff.)}\\
% City, Country \\
% email address or ORCID}
% \and
% \IEEEauthorblockN{6\textsuperscript{th} Given Name Surname}
% \IEEEauthorblockA{\textit{dept. name of organization (of Aff.)} \\
% \textit{name of organization (of Aff.)}\\
% City, Country \\
% email address or ORCID}
}

\maketitle

\begin{abstract}
Laparoscopic surgeries often suffer from reduced visual clarity due to the presence of surgical smoke originated by surgical instruments, which poses significant challenges for both surgeons and vision based computer-assisted technologies. 
In order to remove the surgical smoke, a novel U-Net deep learning with new loss function and integrated differentiable Wiener filter (ULW) method is presented. 
Specifically, the new loss function integrates the pixel, structural, and perceptual properties. Thus, the new loss function, which combines the structural similarity index measure loss, the perceptual loss, as well as the mean squared error loss, is able to enhance the quality and realism of the reconstructed images. 
Furthermore, the learnable Wiener filter is capable of effectively modelling the degradation process caused by the surgical smoke. 
The effectiveness of the proposed ULW method is evaluated using the publicly available paired laparoscopic smoke and smoke-free image dataset, which provides reliable benchmarking and quantitative comparisons. Experimental results show that the proposed ULW method excels in both visual clarity and metric-based evaluation. As a result, the proposed ULW method offers a promising solution for real-time enhancement of laparoscopic imagery. The code is avaiable at \url{https://github.com/chengyuyang-njit/ImageDesmoke}.
\end{abstract}


\section{Introduction}
Surgical smoke is a common byproduct of eletrocautery and laser ablation tools in laparoscopic procedures, and it poses a significant challenge to both human surgeons and automated vision systems. The presence of smoke in the laparoscopic camera view obsures organs and tissues, reducing the surgeons's visual clarity and depth perception\cite{salazar2022removal}. At the same time, this visual degradation can confound computer vision algorithms used for surgical navigation or robotic assistance, often leading to errors in image analysis. The diminished visibility caused by smoke not only increases the cognitive load on surgeons but can also introduce mistakes or delays in image-guided interventions. These factors underscore the importance of effective smoke removal (or “de-smoking”) methods to maintain clear visibility during minimally invasive surgery.

Developing robust smoke removal algorithms has been challenging, partly due to the limitations in available training data. Until recently, there was a lack of paired laparoscopic images with and without smoke - in other words, ground-truth "clear" references for smoky scenes have been scarce\cite{xia2024new}. Consequently, many prior approaches have resorted to indirect means of training and evaluation. Some methods rely on synthetic smoke generation based on physical haze models (treating smoke as a fog-like scattering phenomenon) or simulation in order to create training data. Others have adopted unpaired learning strategies, such as cycle-consistent generative adversarial networks, to translate smoky images to a clear domain without one-to-one supervision\cite{pan2022desmoke}. While these techniques have shown promise, they often suffer from limited generalizability. Models trained on synthetic smoke or unpaired datasets may not capture the full complexity of real in vivo smoke - for example, the dynamic lighting, fluid motion, and particulate characteristics present in actual surgical scenes\cite{xia2024new}. As a result, their performance can degrade when confronted with genuine surgical smoke, which motivated the need for more realistic data and tailored solutions.

A major step toward addressing this problem was the recent introduction of an in vivo paired laparoscopic image dataset for smoke removal\cite{xia2024new}. It is the first dataset to provide real surgical image pairs with and without smoke. It includes 961 paired images from 21 video sequences across 63 prostatectomy procedures. This dataset enables, for the first time, supervised learning and objective evaluation of smoke removal methods on real surgical footage. The study\cite{xia2024new} has also shown that models trained on synthetic or unpaired data struggle on this paired dataset, highlighting the importance of learning from real paired examples.

Deep learning has shown great promise in enhancing diagnostic quality and interpretability\cite{yang2025interpretable} in various medical imaging tasks, such as skin disease detection\cite{inthiyaz2023skin} and public health awareness \cite{yang2024increasing}. Inspired by these advancements, we apply a U-Net-based architecture to tackle visibility challenges in laparoscopic imagery caused by surgical smoke, leveraging the availability of real paired laparoscopic smoke/smoke-free data. Our model integrates a learnable Wiener filter layer to enhance performance by combining data-driven learning with a physics-based prior.

The U-Net backbone\cite{ronneberger2015u}, known for its success in biomedical image segmentation, follows an encoder–decoder structure that captures contextual features and reconstructs a de-smoked image with spatial precision via skip connections. To further refine the output, we embed a differentiable Wiener filter into the network. Traditionally used for deblurring and denoising, the Wiener filter minimizes the mean squared error by estimating and reversing signal degradation under statistical assumptions. By training this filter end-to-end with the rest of the network, it learns to adaptively reduce noise and haze specific to surgical smoke, resulting in sharper and more accurate reconstructions. Recent studies\cite{dong2020deep} show that combining Wiener deconvolution with deep networks improves artifact reduction and detail preservation—motivating its inclusion in our model.

In addition, we employ a multi-objective loss function to guide training. While Mean Squared Error (MSE) ensures pixel-level accuracy, it can result in over-smoothed outputs. To address this, we incorporate Structural Similarity Index (SSIM) loss\cite{zhao2016loss}, which better reflects perceived visual quality by preserving luminance, contrast, and structural integrity. We also include a Perceptual loss\cite{johnson2016perceptual} computed from high-level features of a pre-trained VGG network to enhance texture and realism. Together, these three loss terms balance low-level fidelity with high-level perceptual coherence, enabling the model to produce de-smoked images that are both quantitatively accurate and visually convincing.

Experimental results demonstrate that our approach achieves superior performance in both visual clarity and quantitative metrics, presenting a compelling solution for real-time enhancement of laparoscopic imagery.

\section{Related Work}
Laparoscopic image smoke removal has been explored through a variety of approaches, including traditional image processing, unpaired learning using generative models and paired supervised learning using synthetic smoke data.

Traditional methods treat surgical smoke similarly to fog or haze in out scenes, and transfer outdoor dehazing methods onto laparoscopic image. The most well-known method is dark channel prior (DCP)\cite{he2010single}, which assumes that haze-free images have dark pixels in local patches and estimates transmission maps to recover clearer images. This assumption may work well in outdoor images, but in laparoscopic images, many regions don't contain natural dark pixels, especially under intense endoscopic illumination.

To overcome the lack of paired training data, unpaired image translation methods were introduced. In \cite{sidorov2020generative}, an unsupervised image-to-image translation with a generative adversarial network (GAN) architecture is proposed to remove smoke from laparoscopic surgery images. Likewise, a CycleGan based unpaired method was trained using inter-channel discrepencies and dark channel prior (DCP) \cite{pan2022desmoke}. Authors of \cite{su2023multi} proposed a two-stage method combining pre-trained image translation and CycleGan. However, these methods often suffer from the loss of structural fidelity since the model cannot learn precise pixel-to-pixel correspondences due to lack of aligned smoky and smoke-free image pairs. This can result in distorted anatomy, artifacts, or loss of fine details which are especially critical in medical imaging. Also, since there is no ground truth for direct comparison, it's hard to quantitatively evaluate model performance on real-world images.

Other research has been focused on supervised learning with paired images with synthetic smoke. For example, \cite{salazar2020desmoking} proposed a laparoscopy surgery images desmoking method using an image-to-image translation guided by an embedded dark channel, \cite{song2023vision} used vision transformers for single image dehazing trained on synthetic smoke data. However, synthetic smoke is typically generated using simplified physical models, which often fail to capture the complex behavior of real surgical smoke -- like its texture, density, motion, lighting interaction, and irregular dispersion. This causes a distribution mismatch, leading to poor generalization when the model is applied to real in vivo data. 

The release of the in vivo paired laparoscopic image dataset\cite{xia2024new} marked a turning point in the field, enabling supervised learning on real smoky and smoke-free image pairs. Their study showed that existing methods trained on synthetic or unpaired data underperformed when evaluated on this real dataset, highlighting the importance of training on authentic surgical images.

\begin{figure*}[!t]
\centering
% \framebox[\textwidth]{\parbox{0.9\textwidth}{~\\~\\~\\~\\~\\~\\~\\}}
\includegraphics[width=0.8\textwidth]{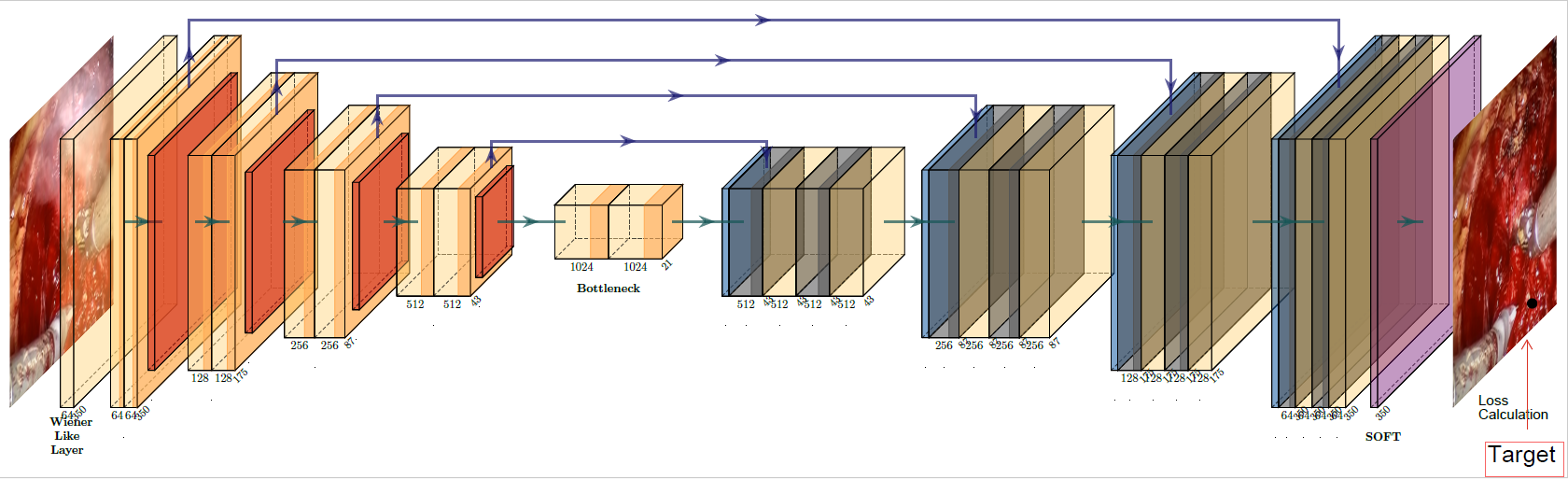}
\caption{The system architecture of the proposed ULW method. The U-Net is used as the backbone while a differentiable Wiener filter layer is integrated after the input of the image with smoke.}
\label{method_architecture}
\end{figure*}

\section{A Novel U-Net Deep Learning with New Loss Function and Integrated Differentiable Wiener Filter (ULW)
method for Laparoscopic Image Desmoking}
We introduce in this section a novel U-Net deep learning with new loss function and integrated differentiable Wiener filter (ULW)
method for surgical smoke removal. 
Specifically, the U-Net deep learning network is used as the backbone architecture and a differentiable Wiener filter layer is integrated into the architecture for effectively modelling the degradation process caused by the surgical smoke. 
Furthermore, a new loss function that integrates the pixel, structural, and perceptual properties is applied. In particular, the new loss funciton combines the Mean Squared Error (MSE) for pixel-level accuracy, and the Structural Similarity Index (SSIM) loss and the perceptual loss for better perceived visual quality and enhanced texture and realism.
The system architecture of the proposed ULW method is shown in Figure \ref{method_architecture}.

\subsection{The U-Net Backbone}
U-Net\cite{ronneberger2015u} is a fully convolutional, symmetric encoder-decoder architecture originally designed for biomedical image segmentation. It consists of a contracting path to capture context and an expansive path that enables precise localization. This model also incorporates skip connections between corresponding layers in the encoder and decoder, which help retain spatial information lost during downsampling. Experiments have shown that it achieves excellent performance in a range of image segmentation tasks, particularly when training data is limited\cite{azad2024medical}. Its flexibility and strong localization capabilities have also made it widely adopted in related areas such as image denoising, super-resolution, and image-to-image translation, and that's the main reason why we adopt U-Net as the backbone for our laparoscopic image desmoking task.

Our backbone consists of 4 levels of downsamping and skip connections, a bottleneck layer with 1024 filters, 4 levels of upsampling using transpose convolutions as is shown in Figure \ref{method_architecture}. 

\subsection{Learnable Wiener Filter}
Wiener filter\cite{gonzales1987digital} is a popular denoising filter. It estimates the original signal from a noisy observation by minimizing the mean squared error(MSE) between the true and estimated signal. It's widely used in image processing, image restoration, communications, etc. Recently, there has been work\cite{dong2020deep} combining Wiener filter with deep learning for non-blind image deblurring and quantitatively outperforms other methods. In our method, we integrate a layer after the input that mimics the function of Wiener filter for the purpose of recovering clean signals from degraded observations. 

Let $\mathbf{x} 
$ be the input smoked image, the Wiener filter's function may be formulated as follows:
\begin{equation}
\hat{\mathbf{x}} = \mathbf{s} \odot \left( \frac{\mathbf{p}}{\mathbf{p} + \boldsymbol{\sigma}_n^2 + \epsilon} \right)\label{eq:wiener}
\end{equation}
where $\mathbf{s} = \mathcal{F}(\mathbf{x})
$ is the filtered signal. The filters are initialized to Gaussian kernels, which mimics the smoothing behavior of classical Wiener filters. $\mathbf{p}=\mathbf{s}^{2}$ is element-wise local signal power estimate while $\boldsymbol{\sigma}_n^2$ is the learnable noise variance. We add a small constant $\epsilon$ in the denominator on the right part of equation\ref{eq:wiener} for numerical stability. The $\odot$ represents element-wise multiplication. 

\subsection{A New Loss Function that Integrates the Pixel, Structural, and Perceptual Properties}

In this section, a new loss function that integrates the pixel, structural, and perceptual properties is presented. In particular, the new loss funciton combines the Mean Squared Error (MSE) for pixel-level accuracy, and the Structural Similarity Index (SSIM) and the perceptual loss for better perceived visual quality and enhanced texture and realism  \cite{pekmezci2024evaluation}, \cite{johnson2016perceptual}.

The SSIM is originally a metric to measure the similarity between two images, focusing on luminance, contrast, and structural information. Unlike traditional losses like Mean Squared Error, which measure pixel-wise differences, SSIM is designed to mimic human visual perception, making it suitable for our image desmoking task, which requires the output image to be of high visual quality for the convenience of surgeons or surgical robots. SSIM compares local patterns of pixel intensities using a sliding window and computes similarity based on luminance (brightness), contrast (variance), structure (correlation). SSIM is a value between 0 and 1, and a higher SSIM indicates better structural similarity. Thus, the opposite of SSIM is used as the SSIM loss. 

The perceptual loss \cite{johnson2016perceptual}, also known as feature reconstruction loss, is a loss function that compares high-level features extracted from a pretrained deep network. Instead of comparing images pixel-by-pixel, perceptual loss computes the difference between the activation maps of intermediate layers of a pretrained model when both input and target images are passed through it. The goal of using perceptual is also to generate images that look good to humans, not just have low pixel-wise error.

Suppose the output of the U-Net backbone is $x_{\text{pred}}$ and the target smoke-free image is $x_{\text{target}}$, function \text{SSIM} outputs the structural similarity between 2 images and function $\phi_l$ outputs the feature map extracted from the intermediate layer of the pre-trained model.
The final loss function can be formulated as follows:

\begin{equation}L = \alpha L_{\text{MSE}}+\beta L_{\text{SSIM}} + \gamma L_{\text{perceptual}}\label{eq:loss}
\end{equation}
\begin{equation}
s.t. \ \ \ \alpha + \beta + \gamma = 1
\end{equation}
where $\alpha$, $\beta$ and $\gamma$ are hyper parameters that can be set accordingly in the experiments and 
\begin{equation}
L_{\text{MSE}} = \left\| x_{\text{pred}} - x_{\text{target}} \right\|^2
\end{equation}

\begin{equation}
L_{\text{SSIM}} = 1 -  \text{SSIM}(x_{\text{pred}},x_{\text{target}})
\end{equation}

\begin{equation}
L_{\text{perceptual}} = \left\| \phi_l(x_{\text{pred}}) - \phi_l(x_{\text{target}}) \right\|^2
\end{equation}

\section{Experiment}
Both subjective and objective evaluations are carried out to comparatively assess the effectiveness of the proposed ULW method for laparoscopic image desmoking. The comparision is among the base model, the foundational model, and the proposed ULW method, and the data set used is the publicly available paired laparoscopic smoke and smoke-free image set that provides reliable benchmarking and quantitative comparisons. The subjective evaluations are implemented by displaying the output smoke free images produced by the base model, the foundational model, and the proposed ULW method, respectively. The objective evaluations apply the SSIM (structural similarity index), PSNR (peak signal to noise ratio), MSE (mean squared error) and CIEDE-2000 \cite{gomez2016comparison} metrics.

\subsection{Experiment Settings}

The first dataset that consists of the paired laparoscopic images with and without smoke\cite{xia2024new} is used to evaluate the performance of the proposed ULW method.  In particular, eighty percent of the 961 paired images are used for training, ten percent of the image pairs are used for validation, and the remaining ten percent of the image pairs are used for testing. 

For the perceptual loss, we use VGG\cite{simonyan2014very} pretrained on ImageNet as the pretrained deep network.

We set $\alpha = \beta = \gamma = \frac{1}{3}$ as the coefficients in the loss functions.

\subsection{Experimental Results for the subjective evaluation}

\begin{figure*}[t]  % or [b] or [h], depending on where you want the figure
    \centering
    \includegraphics[width=0.9\textwidth]{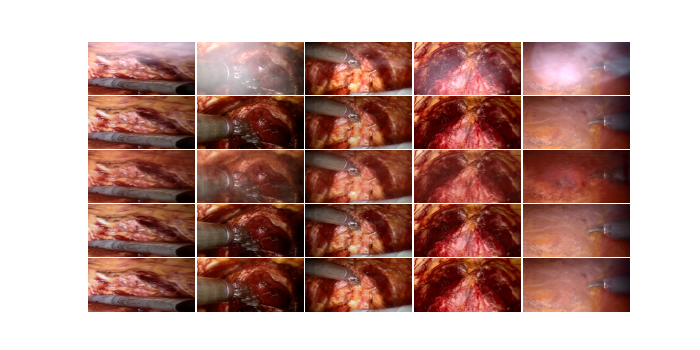}\label{fig:visual}
    \caption{The visual presentation of the desmoking results produced by the base model, the foundational model, and the proposed ULW method, respectively. The first two rows show the paired laparoscopic images with and without smoke, respectively. The third row displays the results of the base model, the fourth row reveals the results of the pix2pix model, and the last row shows the results of the proposed ULW method.}
\end{figure*}

The subjective evaluation results are shown in Figure \ref{fig:visual}. 
Specifically, Figure \ref{fig:visual} displays the visual presentation of the desmoking results produced by the base model, the foundational model, and the proposed ULW method, respectively. In particular, the first two rows show the paired laparoscopic images with and without smoke. The third row displays the results of the base model, the fourth row reveals the results of the pix2pix model, and the last row shows the results of the proposed ULW method.
In figure \ref{fig:visual}, one can see that the smoke has not been completely removed for the image output by the base model (the second example). 
In comparison, the output smoke free images produced by the proposed ULW method achieves the best results.

\subsection{Experimental Results for the objective evaluation}

For the objective evaluation, the metrics SSIM, PSNR, MSE and CIEDE-2000 \cite{gomez2016comparison} are applied to evaluate the quality of the output smoke free images produced by the base model, the foundational model, and the proposed ULW method, respectively.
Note that the CIEDE-2000 metric is the most advanced and perceptually accurate formula developed by the International Commission on Illumination for calculating the color difference between two colors in the CIELAB\cite{gomez2016comparison} color space. 

We compare the proposed ULW method with both the base model and the Pix2Pix model. In the base model, we employ only the U-Net backbone without the learnable Wiener filter and only use the MSE as the loss criterion for training. The Pix2Pix model is based on conditional GAN trained with paired data for pixel-to-pixel translation task. It's a foundational and highly influential baseline in paired image-to-image translation.

Table\ref{tb:res} shows the comparative performance of the base model, the foundational model, and the proposed ULW method using the SSIM, PSNR, MSE, and CIEDE-2000 metrics. 
Using the SSIM metric, the base model achieves 0.9177, the Pix2Pix model achieves 0.9792, and the proposed ULW method achieves the best performance of 0.9907. 
Using the PSNR metric, the base model achieves 21.9394, the Pix2Pix model achieves 30.6189, and the proposed ULW method achieves the best performance of 33.7061. 
Using the MSE metric, the base model achieves 0.0092, the Pix2Pix model achieves 0.0010, and the proposed ULW method achieves the best performance of 0.0006. 
Using the CIEDE-2000 metric, the base model achieves 6.6556, the Pix2Pix model achieves 2.4272, and the proposed ULW method achieves the best performance of 1.8159. 
In conclusion, as we can see from table\ref{tb:res}, in terms of metrics such as the SSIM, the PSNR, the MSE, and the CIEDE-2000, the proposed ULW method not only outperforms the base model by a significant margin, it also surpasses the pix2pix model. 

The superior SSIM score can be attributed to the incorporation of the SSIM loss as an explicit training objective. Furthermore, the integration of a learnable Wiener filter layer effectively suppresses noise, contributing to the highest PSNR and improved perceptual quality. 
These results indicate that our method achieves superior performance in both perceptual and quantitative image quality, demonstrating its effectiveness and generalizability in handling challenging pixel-to-pixel image desmoking tasks.

\begin{table}[!t]
\renewcommand{\arraystretch}{1.3}
\caption{The comparative performance of the base model, the foundational model, and the proposed ULW method using the SSIM, PSNR, MSE, and CIEDE-2000 metrics}\label{tb:res}
\centering
\begin{tabular}{c||c|c|c|c}
\hline
Method & SSIM$\uparrow$ & PSNR$\uparrow$ & MSE$\downarrow$ & CIEDE-2000$\downarrow$\\
\hline\hline
Base Model& 0.9177 &21.9394 & 0.0092 & 6.6556\\
\hline
Pix2Pix\cite{isola2017image} & 0.9792 &30.6189 & 0.0010&2.4272\\
\hline
\textbf{Our Method} &\textbf{0.9907} &\textbf{33.7061} & \textbf{0.0006}&\textbf{1.8159}\\
\hline
\end{tabular}
\end{table}

% \subsection{Ablation Study}
% In this section, we do further experiments to verify the contribution of the proposed components in our method by systematically removing a certain part of the model. 
% \begin{table}[!t]
% \renewcommand{\arraystretch}{1.3}
% \caption{The ablation study : comparative performance of the proposed ULW method w/wo ssim loss, perceptual loss or wiener like layer using the SSIM, PSNR, MSE, and CIEDE-2000 metrics}\label{tb:res}
% \centering
% \begin{tabular}{c||c|c|c|c}
% \hline
% Method & SSIM$\uparrow$ & PSNR$\uparrow$ & MSE$\downarrow$ & CIEDE-2000$\downarrow$\\
% \hline\hline
% w/o wiener& 0.9909 &33.9055 & 0.0007 & 1.8136\\
% \hline
% w/o ssim loss& 0.9675 &29.1446 & 0.0016 & 3.1500\\
% \hline
% w/o perceptual loss & 0.9933&35.8490 & 0.0005&1.3511\\
% \hline
% \textbf{Our Method} &\textb

% \hline
% \end{tabular}
% \end{table}

\section{Conclusion}
A novel U-Net deep learning with new loss function and integrated differentiable Wiener filter (ULW) method for surgical smoke removal is presented in this paper. 
While the differentiable Wiener filter is able to effectively model the degradation process caused by the surgical smoke, the new loss function can integrate the pixel, structural, and perceptual properties. 
Using the real paired laparoscopic smoke dataset, the proposed ULW method demonstrates consistent superiority over both the base U-Net model and the Pix2Pix framework in terms of SSIM, PSNR, MSE, and CIEDE-2000 metrics. 
The integration of the learnable Wiener layer significantly enhances noise suppression and structural fidelity, while the combined loss functions ensure both pixel-level accuracy and perceptual realism. These results validate the effectiveness and generalizability of the proposed approach for improving visibility in minimally invasive surgical procedures.

\bibliographystyle{IEEEtran}
\bibliography{refs}

\end{document}